\documentclass[prd,preprint,nofootinbib]{revtex4}

\usepackage{graphicx}
\usepackage{amssymb}
\usepackage{amsmath}
\usepackage{color}

\begin{document}

\renewcommand{\thefootnote}{\#\arabic{footnote}}
\newcommand{\rem}[1]{{\bf [#1]}}
\newcommand{\gsim}{ \mathop{}_ {\textstyle \sim}^{\textstyle >} }
\newcommand{\lsim}{ \mathop{}_ {\textstyle \sim}^{\textstyle <} }
\newcommand{\vev}[1]{ \left\langle {#1}  \right\rangle }
\newcommand{\bear}{\begin{array}}  
\newcommand {\eear}{\end{array}}
\newcommand{\bea}{\begin{eqnarray}}   
\newcommand{\eea}{\end{eqnarray}}
\newcommand{\beq}{\begin{equation}}   
\newcommand{\eeq}{\end{equation}}
\newcommand{\bef}{\begin{figure}}  
\newcommand {\eef}{\end{figure}}
\newcommand{\bec}{\begin{center}} 
\newcommand {\eec}{\end{center}}
\newcommand{\non}{\nonumber}  
\newcommand {\eqn}[1]{\beq {#1}\eeq}
\newcommand{\la}{\left\langle}  
\newcommand{\ra}{\right\rangle}
\newcommand{\ds}{\displaystyle}
\newcommand{\red}{\textcolor{red}}

\def\SEC#1{Sec.~\ref{#1}}
\def\FIG#1{Fig.~\ref{#1}}
\def\EQ#1{Eq.~(\ref{#1})}
\def\EQS#1{Eqs.~(\ref{#1})}
\def\lrf#1#2{ \left(\frac{#1}{#2}\right)}
\def\lrfp#1#2#3{ \left(\frac{#1}{#2} \right)^{#3}}
\def\GEV#1{10^{#1}{\rm\,GeV}}
\def\MEV#1{10^{#1}{\rm\,MeV}}
\def\KEV#1{10^{#1}{\rm\,keV}}

\def\REF#1{(\ref{#1})}
\def\lrf#1#2{ \left(\frac{#1}{#2}\right)}
\def\lrfp#1#2#3{ \left(\frac{#1}{#2} \right)^{#3}}

\newcommand{\nuR}{\tilde{\nu}_R}

\begin{flushright}
\end{flushright}

\title{
Cosmic rays from Leptonic Dark Matter
}

\author{Chuan-Ren Chen and  Fuminobu Takahashi}

\affiliation{Institute for the Physics and Mathematics of the Universe,
University of Tokyo, Chiba 277-8568, Japan }


\begin{abstract}
  If dark matter possesses a lepton number, it is natural to expect
  the dark-matter annihilation and/or decay mainly produces the
  standard model leptons, while negligible amount of the antiproton
  is produced.  To illustrate such a simple idea, we consider a
  scenario that a right-handed sneutrino dark matter decays into the
  standard model particles through tiny $R$-parity violating
  interactions. Interestingly enough, charged leptons as well as
  neutrinos are directly produced, and they can lead to a sharp peak
  in the predicted positron fraction. Moreover, the decay of the
  right-handed sneutrino also generates diffuse continuum gamma rays
  which may account for the excess observed by EGRET, while the
  primary antiproton flux can be suppressed. Those predictions on the
  cosmic-ray fluxes of the positrons, gamma rays and antiprotons will
  be tested by the PAMELA and FGST observatories.
\end{abstract}

\preprint{IPMU 08-0071}
\pacs{98.80.Cq}

\maketitle

\section{Introduction}
\label{sec:1}
The presence of dark matter has been securely established by numerous
observational evidences. In particular, the latest 5yr WMAP data
determined the dark matter abundance with unprecedented precision
as~\cite{Komatsu:2008hk}
\beq
\Omega_{\rm DM} h^2 \;=\;0.1099 \pm 0.0062,
\label{eq:dm-abundance}
\eeq
where $\Omega_{\rm DM}$ is the fraction of critical density in dark
matter and $h$ is the Hubble parameter in units of $100$
km/Mpc/sec. In the meantime it is not known yet what dark matter is
made of despite many experimental direct/indirect searches hitherto
$-$ but there is a hope that the PAMELA~\cite{Picozza:2006nm} and FGST
(formerly GLAST)~\cite{FGST} satellites may provide us with some
important information on the nature of dark matter.

It is normally assumed that the dark matter is charged under an exact
discrete symmetry to ensure its stability, for instance, $R$-parity in
a supersymmetric theory. There might exist many discrete symmetries
realized in our vacuum and one of them may be responsible for making
the dark matter stable. However, if the discrete symmetry is broken,
the dark matter will be unstable and eventually decay into the
Standard Model (SM) particles. Recently such a decaying dark matter
has attracted much attention (e.g. the gravitino with $R$-parity
violation~\cite{Takayama:2000uz,Buchmuller:2007ui,Bertone:2007aw,
  Ibarra:2007wg,Ishiwata:2008cu} or a hidden $U(1)$ gauge
boson~\cite{Chen:2008yi} and gaugino~\cite{Ibarra:2008kn}), since the
high-energy cosmic rays produced by the decay of dark matter may
account for the observed anomalous excesses in gamma
rays~\cite{Sreekumar:1997un,Strong:2004ry} and/or positrons~\cite{Barwick:1997ig}.

Very recently, the PAMELA data on the antiproton flux has been
released~\cite{Boezio:2008mp}, and it suggests that the observed
antiprotons are mainly secondaries, which is consistent with the
balloon-borne experiments at the top of atmosphere
(TOA)~\cite{Orito:1999re}. This will place tight constraints on
possible dark matter candidates for explaining the anomalous excesses
in positrons and/or gamma rays~\cite{Barger:2008su}.  Of course, the
predicted antiproton flux has a large uncertainty mainly due to our
poor understanding of the cosmic-ray propagation inside our
Galaxy. However, if we take the constraint on the antiproton flux
seriously, and if we attribute the excesses in positrons and gamma
rays to the dark-matter annihilation or decay, we are led to
explore a dark matter candidate naturally satisfying the constraint on the antiproton flux. 
Our main idea is as follows.  If the dark
matter has a lepton number, it is quite natural to expect that it
annihilates or decays mainly into the leptons, while only negligible
amount of the antiprotons are produced.  The purpose of this paper is
to illustrate this simple and naive idea by using an explicit example.

There are several candidates for leptonic dark matter such as the
sterile neutrino~\cite{Dodelson:1993je}.  As an example, in this paper
we consider a scenario that a right-handed sneutrino $\nuR$ is the
lightest supersymmetric particle (LSP) and accounts for dark matter of
the universe.  The neutrino mass is assumed to be Dirac type in our
study. If the $R$-parity is an exact symmetry of nature, the $\nuR$
dark matter is absolutely stable~\cite{Asaka:2005cn,Gopalakrishna:2006kr,Lee:2007mt}.  
However, the $R$-parity may not be an exact symmetry and is only an approximate one
accompanied with tiny violations. We focus on the $R$-parity violating
bilinear term throughout this paper. In the absence of the $R$-parity,
the $\nuR$ dark matter is not stable anymore, and directly decays into
the charged leptons ($\tau$, $\mu$ and $e$) and neutrinos, and it can
also decay into the quarks and the $W$ and $Z$ gauge bosons depending
on the mixing with the Higgs bosons.  As a result, a sharp peak is
predicted in the positron fraction, which may explain the anomalous
excess observed by High Energy Antimatter Telescope
(HEAT)~\cite{Barwick:1997ig}, MASS~\cite{Grimani:2002yz} and
AMS~\cite{Aguilar:2007yf} experiments. Interestingly, the preliminary
PAMELA data also exhibits such anomalous excess in the cosmic-ray
positron fraction, although we do not try to include the preliminary
data since it is not formally released yet.  The continuum gamma rays
are also produced mainly from the decay of pions, and those gamma rays
may account for the excesses observed by
EGRET~\cite{Sreekumar:1997un}. For a proper set of parameters, the
production of antiprotons can be suppressed, which is consistent with
an observational fact that the antiprotons measured by the
balloon-borne experiments~\cite{Orito:1999re} and also by the PAMELA
satellite~\cite{Boezio:2008mp} are considered to be mainly
secondaries~\cite{Mitchell:1996bi}.  The suppression in the
antiproton flux is particularly important because some decaying dark
matter scenarios predict too large antiproton flux at the solar
system~\cite{Ibarra:2007wg}.

This paper is organized as follows. In Sec.\ \ref{sec:model}, we
present the effective interactions and the relevant decay modes used
in our calculations.  In Sec.\ \ref{sec:numerical}, we calculate the
spectra of the positron, gamma-ray, and antiproton produced from the
decay of the $\nuR$, and compare them with the observational data.  We
also discuss how the right-handed sneutrinos are produced in the early
universe in Sec.\ \ref{sec:production}, and we give our discussions
and conclusions in Sec.\ \ref{sec:Conclusion}.

\section{Framework}
\label{sec:model} 
The non-vanishing neutrino masses have been firmly established by
neutrino oscillation experiments (see Ref.~\cite{GonzalezGarcia:2007ib} for
recent review and references therein), although it is not known yet
whether the neutrinos are Majorana or Dirac fermions. In this paper,
we consider a Dirac-type neutrino mass in a supersymmetry theory without
$R$-parity conservation, and assume the gravity mediation.  We introduce two additional interactions to
the minimal supersymmetric standard model (MSSM); one is the neutrino
Yukawa coupling for the Dirac neutrino mass, and the other is the
$R$-parity violating bilinear term.  The superpotential is
therefore~\footnote{ The right-handed sneutrinos acquire tiny vacuum
  expectation values (vev) after the electroweak symmetry breaking in
  the presence of the bilinear $R$-parity violating term. However, the
  expectation values are suppressed by the neutrino Yukawa couplings,
  and so, we can neglect their effects. On the other hand, the
  bilinear term can be induced from the neutrino Yukawa couplings with
  non-vanishing vevs of the right-handed sneutrinos or the linear
  terms in the K\"ahler potential.  }
\bea W &=& W_{\rm MSSM} + y_{ij}^{\nu} \bar{N}_{i} L_{j}H_{u} +
\mu_{i}H_{u}L_{i},\\
\label{eq:Yukawa}
W_{\rm MSSM} &=& y^u_i \bar{U}_i Q_i H_u - y^d_{i} \bar{D}_i Q_i H_d -
y^e_i \bar{E}_i L_i H_d + \mu H_u H_d, \eea
where $\bar{N}$ is the right-handed neutrino superfield, $y^{\nu}$ is
the neutrino Yukawa coupling, $\mu_i$ denotes the coefficient of the
$R$-parity violating bilinear term, and the indices $i$ and $j$ denote
the generations.  We neglect the flavor mixings in the MSSM Yukawa
interactions, and assume the minimal K\"ahler potential for all the
MSSM particles as well as the right-handed neutrinos.

Let us first discuss the neutrino Yukawa coupling.  The three
left-handed neutrinos in the weak eigenstate $\nu_i$ are related with
the mass eigenstates $\hat{\nu}_i$ as
\begin{equation}
\nu_{i}\;=\;U_{ij} \cdot \hat{\nu}_{j},
\label{eq:Ngauge_Nmass}
\end{equation}
where $i$ runs from $1$ to $3$ and the well-known mixing matrix $U$ takes the
following form:
\begin{equation}
U\;=\;\left(\begin{array}{ccc} c_{13}c_{12} & s_{12}c_{13} &
  s_{13}\\ -s_{12}c_{23}-s_{23}s_{13}c_{12} &
  c_{23}c_{12}-s_{23}s_{13}s_{12} &
  s_{23}c_{13}\\ s_{23}s_{12}-s_{13}c_{23}c_{12} &
  -s_{23}c_{12}-s_{13}s_{12}c_{23} & c_{23}c_{13}\end{array}\right),
\label{eq:Neu_mixing}
\end{equation}
where $c_{ij}\equiv\cos\theta_{ij}$ and $s_{ij}\equiv\sin\theta_{ij}$,
and $\theta_{ij}$ denotes the mixing angle of neutrinos $\nu_i$ and
$\nu_j$.
We can define the right-handed neutrinos so that the neutrino Yukawa
coupling matrix is given by
\begin{equation}
y^{\nu} \;=\; {\rm diag}(m_{1},m_{2},m_{3}) U^\dag,
\label{eq:redef_Yukawa}
\end{equation}
where $m_i$ is the mass of $\hat{\nu}_i$. In
Table~\ref{tab:neutrino_exp}, we show the observational constraints on
those mixings and the mass differences based on the global
three-neutrino analysis \cite{Schwetz:2008er}. Since the neutrino
oscillation data is not sensitive to the absolute masses of the
neutrinos, the following three neutrino mass spectra are possible: (i)
normal hierarchy case $m_{3}>m_{2}\gg m_{1}$ (ii) inverted hierarchy
case $m_{2}>m_{1}>m_{3}$ and (iii) degenerate case $m_{1}\simeq
m_{2}\simeq m_{3}$.  Throughout this paper, we adopt the normal
hierarchy for simplicity, therefore, $m_{2}^{2}\sim\Delta m_{21}^{2}$,
$m_{3}^{2}\sim\Delta m_{31}^{2}$ and we adopt massless $m_1$ in our
numerical study.
\begin{table}
\begin{tabular}{|c|c|c|c|c|c|}
\hline 
&
$\Delta m_{21}^{2}\,[{\rm eV}^{2}]$&
$|\Delta m_{31}^{2}|\,[{\rm eV}^{2}]$&
$\sin^{2}\theta_{12}$&
$\sin^{2}\theta_{23}$&
$\sin^{2}\theta_{13}$\tabularnewline
\hline
\hline 
best fit&
$7.65_{-0.20}^{+0.23}\times10^{-5}$&
$2.40_{-0.11}^{+0.12}\times10^{-3}$&
$0.304_{-0.016}^{+0.022}$&
$0.50_{-0.06}^{+0.07}$&
$0.01_{-0.011}^{+0.016}$\tabularnewline
\hline
adopted&
$7.65\times10^{-5}$&
$2.4\times10^{-3}$&
$0.304$&
$0.50$&
$0$\tabularnewline
\hline
\end{tabular}

\caption{The best-fit and adopted values of three-flavor neutrino oscillation parameters
from global data, including solar, atmospheric, reactor (KamLAND,
CHOOZ) and accelerator (K2K) experiments~\cite{Schwetz:2008er}. }
\label{tab:neutrino_exp}
\end{table}

Next we consider the $R$-parity violation.  In the presence of the
bilinear $R$-parity violation, it is known that there is no unique way
to define $L_{i}$ and $H_{d}$  since they have
the same quantum numbers~\cite{Barbier:2004ez}.  Taking account of the soft SUSY breaking
terms, the left-handed sneutrino $\la L_i^0 \ra$ acquires a
non-vanishing vacuum expection value (vev) after the electroweak
breaking in a general basis.  In our work, we adopt a basis such that
the $\la L_i^0 \ra$ vanishes by a proper redefinition of $L_i$ and
$H_d$.  In this basis, the trilinear $R$-parity violating interactions
are generically induced. Furthermore, the right-handed sneutrinos get
mixed with the left-handed sneutrinos as well as up- and down-type
Higgs, which would make analysis on the right-handed sneutrino decay
complicated.  The purpose of this paper is not to explore all possible
parameter space, but to illustrate our basic idea that the dark matter
with a lepton number can account for the sharp rise in the positron
fraction while the antiproton flux can be suppressed. Thus we focus
on a case that the dark matter is comprised of $N_3$, which decays
into the SM particles through the $R$-parity violating bilinear term
with $\mu_1 \ne 0$.  Then, the relevant mixings of $N_3$ with the
left-handed sneutrinos and the Higgs bosons are suppressed by the small
mixing angle $\theta_{13}$. In particular, those mixings are absent in
the limit of $\theta_{13}=0$, which therefore greatly simplifies our
analysis.  The adopted values of the mass differences and the mixing
angles are shown in Table~\ref{tab:neutrino_exp}.  We will discuss
later how our result is modified for other choices of the parameters.

\begin{figure}
\includegraphics[scale=0.5]{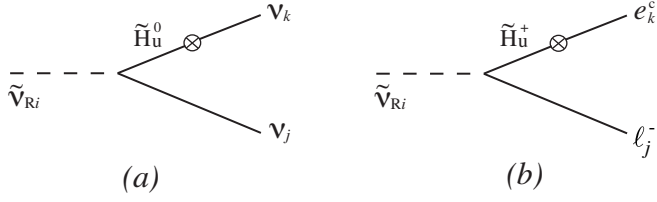}
\caption{Feynman diagrams of right-handed sneutrino decay via the mixing between
the up-type higgsino and the SM leptons.\ \label{fig:huL_mixing}}
\end{figure}

In our set-up, the main decay channels of the right-handed sneutrino,
$\tilde{\nu}_{R3}^c$ (the scalar component of $\bar N_3$), are
neutrinos and charged leptons through the neutrino-neutralino and
charged-lepton-chargino mixings.  The corresponding Feynman diagrams
are shown in Fig.~\ref{fig:huL_mixing}.  We can see from
Eq.~(\ref{eq:Yukawa}) that the interactions between the right-handed
sneutrino $\nuR$ with the higgsino and the SM leptons are given by
\begin{equation}
{\cal L} \;\supset\;-y_{ij}^{\nu}\tilde{\nu}_{Ri}^c \ell_{j}^{-}\tilde{H}_{u}^{+}+y_{ij}^{\nu} \tilde{\nu}_{Ri}^c \nu_{j}\tilde{H}_{u}^{0},\,
\label{eq:RSN-Hu-L}
\end{equation}
 where $\tilde{\nu}_{Ri}^c$ denotes the scalar component of
 $\bar{N}_i$, and $\ell_{1}=e,\,\ell_{2}=\mu$ and $\ell_{3}=\tau$.
 First let us consider the neutrino-neutralino mixing. The mass matrix
 of the neutral fermions $M_{N}$ is given by ${\cal L} \supset
 (-1/2)\Psi^{0T}M_{N}\Psi^{0}$ with~\cite{Barbier:2004ez}
\bea
M_{N}&=&\left(\begin{array}{cc}
M^{(N)}_{MSSM} & M^{(N)}_{\not R}\\
M^{(N) T}_{\not R} & 0\end{array}\right),\\
\label{eq:neutralHu-N}
\eea
where we have defined $\Psi^{0} \equiv
(\tilde{B},\,\tilde{W}^{3},\,\tilde{H}_{d}^{0},\,\tilde{H}_{u}^{0},\nu_{Li})^T$,
$M^{(N)}_{MSSM} $ is the usual neutralino mass matrix in MSSM, and
$M^{(N)}_{\not R}$ is the mass matrix between the neutralinos and the
left-handed neutrinos arising from the $R-$parity violating
interactions. The explicit expressions of $M^{(N)}_{MSSM} $ and
$M^{(N)}_{\not R}$ are
\begin{equation}
M^{(N)}_{MSSM} =\left(\begin{array}{cccc}
M_{1} & 0 & -m_{Z}s_{W}c_{\beta} & m_{Z}s_{W}s_{\beta}\\
0 & M_{2} & m_{Z}c_{W}c_{\beta} & -m_{Z}c_{W}s_{\beta}\\
-m_{Z}s_{W}c_{\beta} & m_{Z}c_{W}c_{\beta} & 0 & -\mu\\
m_{Z}s_{W}s_{\beta} & -m_{Z}c_{W}s_{\beta} & -\mu & 0\end{array}\right),\quad 
M^{(N)}_{\not R}=\left(\begin{array}{ccc}
 & 0_{3\times3}\\
-\mu_{1} & -\mu_{2} & -\mu_{3}\end{array}\right),\,
\label{eq:Hu-N_matrix}
\end{equation}
where $c_{W}\equiv\cos\theta_{W}$ and $s_W\equiv\sin\theta_{W}$ with
$\theta_{W}$ being the weak mixing angle; $c_{\beta}\equiv\cos\beta$
and $s_{\beta}\equiv\sin\beta$ with $\tan\beta = v_{u}/v_{d}$ where
$v_{u} (v_{d})$ is the vev of up-type (down-type) Higgs field.  Here
we keep $\mu_2$ and $\mu_3$ in the mass matrix, although we will set
them to be zero in the following analysis.  The mass matrix $M_{N}$
could be diagonalized by an unitary matrix $V$, and the mixing between
the up-type higgsino $\tilde{H}_{u}^{0}$ and the left-handed neutrinos
$\nu_{Li}$ is given by $V_{4,i+4}$. In a similar way, the charged
leptons mix with the charginos via the $R$-parity violating
interactions. The mass matrix $M_{C}$ of the charginos and the charged
leptons is given by ${\cal L} \supset -\Psi^{-T}M_{C}\Psi^{+}$ with~\cite{Barbier:2004ez}
\begin{equation}
M_{C}=\left(\begin{array}{cc}
M^{(C)}_{MSSM} & 0_{2\times3}\\
M^{(C)}_{\not R} & y^{e}_i \delta_{ij} v_{d}/\sqrt{2}\end{array}\right),\,
\label{eq:chargeHu_L}
\end{equation}
where we have defined $\Psi^{-} \equiv
(\tilde{W}^{-},\,\tilde{H}_{d}^{-}, \ell_i)^T$ and $\Psi^{+} \equiv
(\tilde{W}^{+},\,\tilde{H}_{u}^{+},e^c_i)^T$, $M^{(C)}_{MSSM}$ is the
usual chargino mass matrix in MSSM, 
$M^{(C)}_{\not R}$ is the mass matrix induced by $R$-parity violating interactions
and $y^{e}$ is the charged lepton Yukawa coupling. The explicit
expressions are
\begin{equation}
M^{(C)}_{MSSM}\;=\;\left(\begin{array}{cc}
M_{2} & \sqrt{2}m_{Z}c_{W}s_{\beta}\\
\sqrt{2}m_{Z}c_{W}c_{\beta} & \mu\end{array}\right),\quad
M^{(C)}_{\not R}\;=\;\left(\begin{array}{cc}
0 & -\mu_{1}\\
0 & -\mu_{2}\\
0 & -\mu_{3}\end{array}\right)
\label{eq:Hup-L_mixing}
\end{equation}
The mass matrix $M_{C}$ can be diagonalized by two rotational matrices
$O_{-}$ and $O_{+}$ which relate the gauge eigenstates of $\Psi^{-}$ and
$\Psi^{+}$ with their mass eigenstates, respectively,
i.e. $O_{-}^{T}M_{C}O_{+}={\rm
  diag}(m_{\tilde{\chi}_{1}^{-}},m_{\tilde{\chi}_{2}^{-}},m_{e},m_{\mu},m_{\tau})$.
The mixings between $\tilde{H}_{u}^{+}$ and the right-handed charged
leptons $e^{c+}_{i}$ are given by the elements
$\left[O_{+}\right]_{2,2+i}$.  Combining the
interaction~(\ref{eq:RSN-Hu-L}) and the mixings between the up-type
higgsino and the SM leptons, the $\nuR$ will decay into the SM leptons
as shown in Fig.~\ref{fig:huL_mixing}.

Finally, we summarize the relevant decay processes for
our study below;
\begin{eqnarray}
\tilde{\nu}_{Ri} &\to&\nu_{j} \bar\nu_{k}  \propto  |y_{ij}^{\nu} [V]_{4,k+4}|^{2},\nonumber \\
\tilde{\nu}_{Ri}&\to&\ell_{j}^{-} e_{k}^{c+}  \propto  |y_{ij}^{\nu}\left[O_{+}\right]_{2,2+k}|^{2},
\label{eq:RSN_decay}
\end{eqnarray}
 where $V$ and $O_{\pm}$ 
 are the rotation matrices in the neutralino-neutrino and the
chargino-lepton
respectively.
In Table~\ref{tab:Brs}, we show the branching ratios of a decaying $\tilde{\nu}_{R3}$, in which we assume non-vanishing  $\mu_{1}$ with
$\mu_{2,3}=0$ for simplicity. 
Notice that positrons are always produced in the charged lepton decay modes since only $\mu_1$
is non-vanishing. Also, because of the smallness of the positron mass, 
the decay branching ratios of positron are highly suppressed down to a few percentages,
compared to that of the neutrino production.

\begin{table}
\begin{tabular}{|c|c|c|c|}
\hline 
channel&
$e^{+}\mu^{-}$&
$e^{+}\tau^{-}$&
$\nu\bar{\nu}$\tabularnewline
\hline 
branching ratios (\%)&
$1.5$&
$1.5$&
$97$\tabularnewline
\hline
\end{tabular}
\caption{Branching ratios of right-handed sneutrino $\tilde{\nu}_{R3}$.}
\label{tab:Brs}
\end{table}

\section{Cosmic-ray fluxes}
\label{sec:numerical} 

Let us here mention that both the right-handed sneutrino and its
antiparticle can be dark matter and contribute to the cosmic-ray
signals. For simplicity, we assume that both of them have been
produced with an equal amount in the early universe, therefore, we
have
\begin{equation}
\Omega_{DM}\equiv\frac{\rho_{\nuR}+\rho_{\tilde{\nu}_{R}^{c}}}{\rho_{c}}=2\frac{\rho_{\nuR}}{\rho_{c}}=2\Omega_{\nuR},
\label{eq:DM}
\end{equation}
where $\Omega_{DM}$ and $\Omega_{\nuR}$ are density parameters of the
dark matter and the right-handed sneutrino, respectively; $\rho_{c}$
is the critical density and $\rho_{\nuR (\tilde{\nu}_{R}^{c})}$ is the
energy density of right-handed (anti)sneutrino.  In the rest of this
paper, contributions of both right-handed sneutrino and its
antiparticle are included in the cosmic-ray fluxes.  The method for
calculations of the gamma-ray flux and positron fraction are the same
as that in our previous study~\cite{Chen:2008yi}, therefore, we only
show the equations which are needed in the calculations. We refer
readers who are interested in the derivations to
Ref.~\cite{Ibarra:2007wg} and references therein.

\begin{figure}
\includegraphics[scale=0.6]{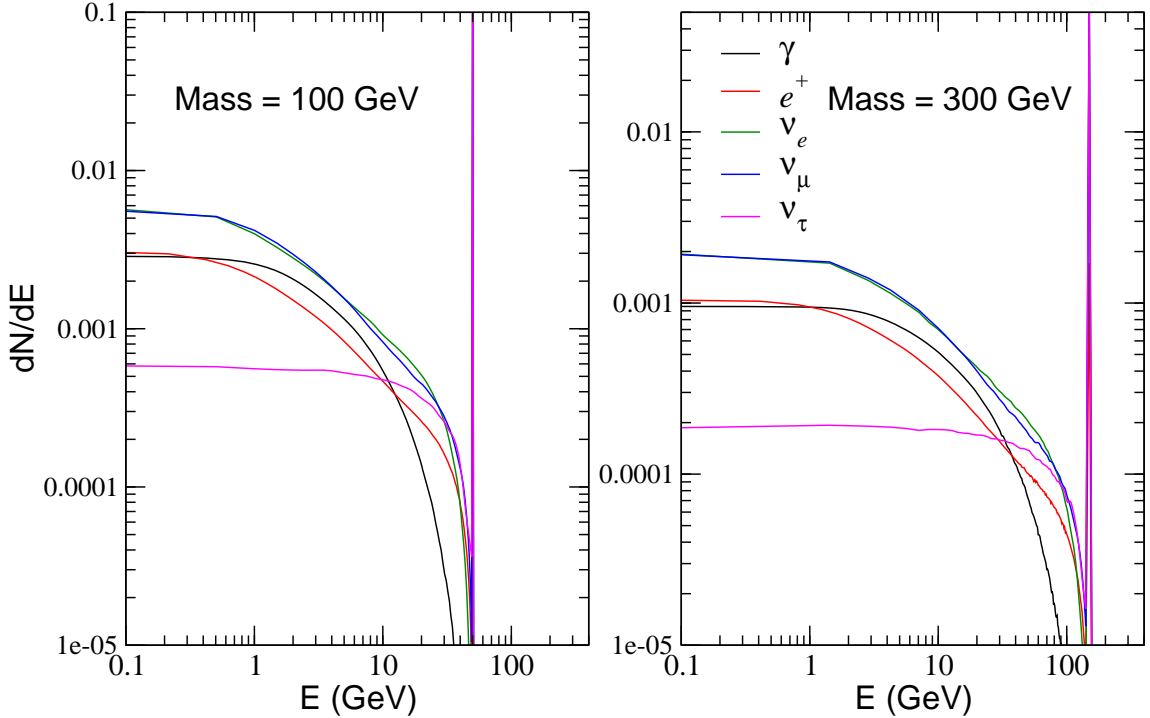}
\caption{Energy spectra of gamma, positron and neutrinos generated from the decay
of a $\nuR$.}
\label{fig:dnde}
\end{figure}

\subsection{Positron fraction}
After being produced from the decay of the right-handed sneutrino,
positrons will propagate in the magnetic field of the Milky Way. The
positron flux is given by
\begin{equation}
\Phi_{e^{+}}^{prim}(E)=\frac{c}{4\pi
m_{\nuR}\tau_{\nuR}}\int_{0}^{m_{\nuR}/2}dE'G(E,E')\frac{dN_{e^{+}}}{dE'},
\label{eq:flux_e} 
\end{equation} 
where $E$ is in units of GeV, and
$m_{\nuR}$ and $\tau_{\nuR}$ are the mass and lifetime of the
right-handed sneutrino dark matter. Here $dN_{e^{+}}/dE$ is the energy spectrum
of positron from the dark matter decay. We have used 
PYTHIA\ \cite{Sjostrand:2006za} to calculate the energy spectra, and the numerical
results are shown in
Fig.\ \ref{fig:dnde}.
The Green function $G(E,E')$ in Eq.~(\ref{eq:flux_e}) is approximately given
by~\cite{Ibarra:2007wg} 
\begin{equation}
G(E,E')\simeq\frac{10^{16}}{E^{2}}e^{a+b(E^{\delta-1}-E'^{\delta-1})}\theta(E'-E)\quad [{\rm
sec/cm}^{3}],
\label{eq:G_approx} 
\end{equation} 
where $\delta$ is related
to the properties of the interstellar medium and can be determined mainly
from the  Boron to Carbon ratio (B/C)\ \cite{Maurin:2001sj}.
In the numerical study, we adopt the following parameters,
$\delta=0.55$, $a=-0.9716$ and $b=-10.012$\ \cite{Ibarra:2007wg},
that are consistent with the B/C value and produce the minimum flux
of positrons.

\begin{figure}
\includegraphics[scale=0.5]{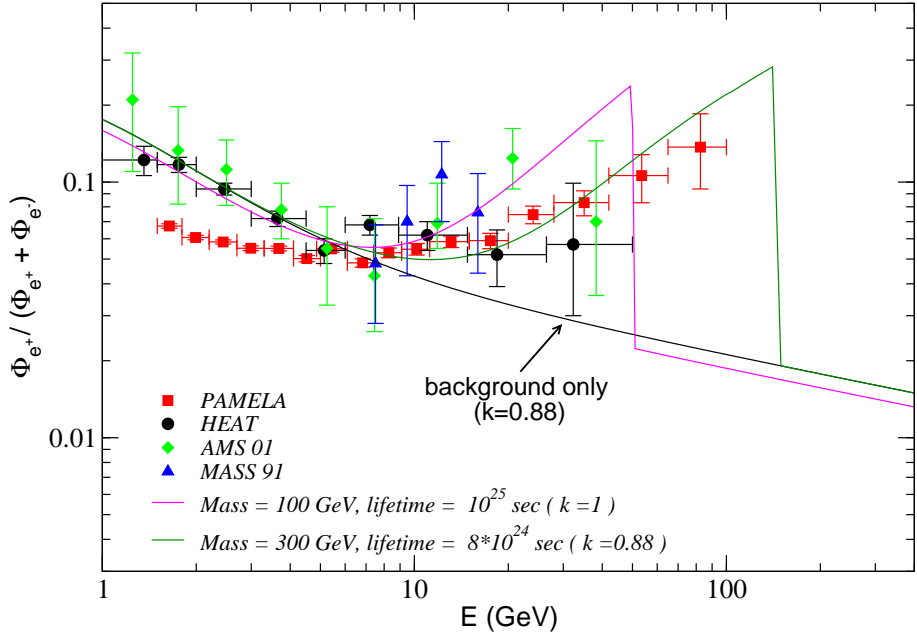}
\caption{The fraction of positron flux from right-handed sneutrino dark matter decay, shown
together with experimental data~\cite{Barwick:1997ig,Grimani:2002yz,Aguilar:2007yf,Adriani:2008zr}.}
\label{fig:Fraction_e}
\end{figure}

In addition to the positron flux from dark matter decay, there exist
secondary positrons produced from interactions between the primary cosmic rays and
nuclei in the interstellar medium. The positron flux is considered to
suffer from the solar modulation, especially for the energy below
$10$ GeV. If the solar modulation effect is independent of the
charge-sign, one can cancel the effect by taking the positron
fraction,
\begin{equation}
\frac{\Phi_{e^{+}}}{\Phi_{e^{+}}+\Phi_{e^{-}}},
\label{eq:ep_exp}
\end{equation}
which is indeed measured in many experiments. To estimate the positron fraction, it is
necessary to include the electron flux. We use the approximations of
the $e^{-}$ and $e^{+}$ background fluxes~\cite{Moskalenko:1997gh,Baltz:1998xv} 
\bea
\Phi_{e^{-}}^{prim}(E) &=&\frac{0.16E^{-1.1}}{1+11E^{0.9}+3.2E^{2.15}}\quad[{\rm GeV}^{-1}{\rm cm}^{-2}{\rm sec}^{-1}{\rm sr}^{-1}],\nonumber \\
\Phi_{e^{-}}^{sec}(E) & =&\frac{0.7E^{0.7}}{1+110E^{1.5}+600E^{2.9}+580E^{4.2}}\quad[{\rm GeV}^{-1}{\rm cm}^{-2}{\rm sec}^{-1}{\rm sr}^{-1}],\nonumber \\
\Phi_{e^{+}}^{sec}(E)&=&\frac{4.5E^{0.7}}{1+650E^{2.3}+1500E^{4.2}}\quad[{\rm GeV}^{-1}{\rm cm}^{-2}{\rm sec}^{-1}{\rm sr}^{-1}],
\label{eq:bg_e}
\eea
where $E$ is in units of GeV. Therefore, the fraction of $e^{+}$
flux is 
\begin{equation}
\frac{\Phi_{e+}^{prim}+\Phi_{e+}^{sec}}{\Phi_{e+}^{prim}+\Phi_{e+}^{sec}+k\Phi_{e^{-}}^{prim}+\Phi_{e^{-}}^{sec}},
\label{eq:fract_e}
\end{equation}
where $k$ is a free parameter which is used to fit the data when no
primary source of $e^{+}$ flux
exists~\cite{Baltz:1998xv,Baltz:2001ir}.  Our numerical results for
$m_{\nuR}=100$ GeV and $300$ GeV are shown in
Fig.~\ref{fig:Fraction_e}.  The typical feature is the sharp turnup at
$E\approx10$ GeV and a drop-off at $E=m_{\nuR}/2$ due to the
contribution of the direct production of $e^{+}$ from the decay of
dark matter~\cite{Chen:2008yi}.

\subsection{Gamma-ray flux}
Generally, the main contribution to the gamma-ray spectrum arises from
the pion $\pi^{0}$ generated in the QCD hadronization process of
$q\bar{q}$ and in the decay of $\tau$. Since no quark pair is produced
in our set-up, the only source of $\pi^0$ is from the decay of
$\tau$. To estimate the spectrum, we use the
PYTHIA\ \cite{Sjostrand:2006za} Monte Carlo program with the branching
ratios shown in Table~\ref{tab:Brs} and the energy spectra
$dN_{\gamma}/dE$ are presented in Fig.~\ref{fig:dnde}.  It is worth
noting that there is no line emission of the gamma rays from the decay
of $\nuR$, which is present in the case of the gravitino dark
matter~\cite{Ibarra:2007wg,Ishiwata:2008cu}.

There are galactic and extragalactic contributions from the decay of
$\nuR$ to the observed gamma-ray flux. The flux of the gamma-ray from
the extragalactic origin is estimated
as\ \cite{Ibarra:2007wg,Ishiwata:2008cu}
\begin{equation}
\left[E^{2}\frac{dJ_{\gamma}}{dE}\right]_{eg}=\frac{E^{2}c\Omega_{DM}\rho_{c}}{4\pi m_{\nuR}\tau_{\nuR}H_{0}\Omega_{M}^{1/2}}\int_{1}^{y_{{\rm eq}}}dy\frac{dN_{\gamma}}{d(yE)}\frac{y^{-3/2}}{\sqrt{1+\frac{\Omega_{\Lambda}}{\Omega_{M}}y^{-3}}},
\label{eq:gamma_eg}
\end{equation}
where $c$ is the speed of light; $\Omega_{M}$ and $\Omega_{\Lambda}$
are the density parameters of matter (including both baryons and dark
matter) and the cosmological constant, respectively; $H_{0}$ is the
Hubble parameter at the present time; $y\equiv1+z$, where $z$ is the
redshift, and $y_{{\rm eq}}$ denotes a value of $y$ at the
matter-radiation equality. For the numerical results, we
use\ \cite{Komatsu:2008hk}
\begin{equation}
\Omega_{DM}h^{2}=0.1099,\quad\Omega_{M}h^{2}=0.1326,\quad\Omega_{\Lambda}=0.742,\quad\rho_{c}=1.0537\times10^{-5}{\rm GeV/cm}^{3}.\,
\label{eq:input}
\end{equation}

\begin{figure}
\includegraphics[scale=0.5]{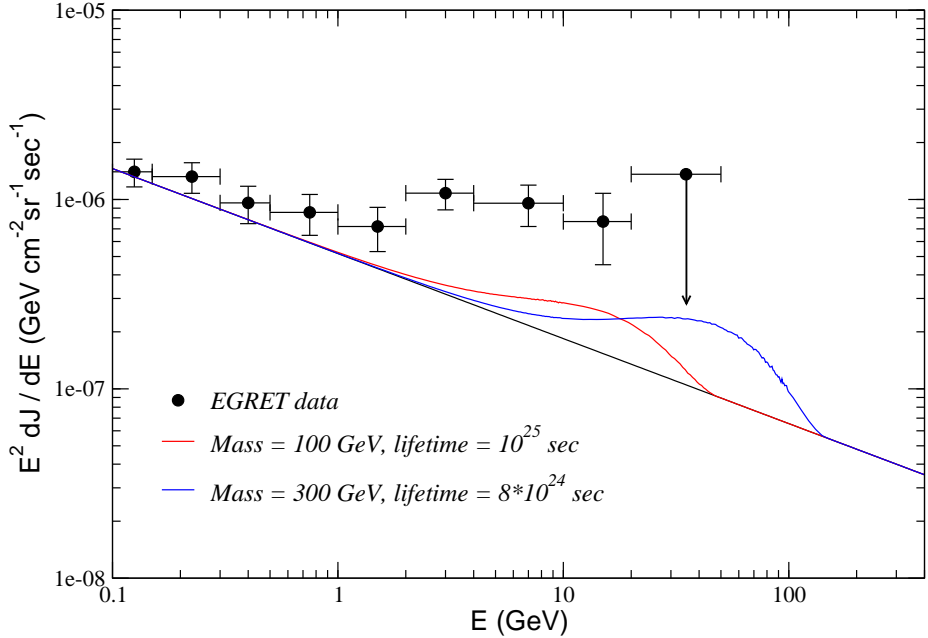}
\caption{Gamma-ray flux predicted from decay of right-handed sneutrino dark matter,
shown together with the EGRET data~\cite{Sreekumar:1997un}.}
\label{fig:gamma}
\end{figure}

On the other hand, the gamma-ray flux from the decay of dark matter in
the Milky Way halo is
\begin{equation}
\left[E^{2}\frac{dJ_{\gamma}}{dE}\right]_{halo}=\frac{E^{2}}{4\pi m_{\nuR}\tau_{\nuR}}
\frac{dN_{\gamma}}{dE}\left\langle \int_{los}\rho_{halo}(\vec{\ell})d\vec{\ell}\right\rangle,
\label{eq:gamma_halo}\end{equation}
where $\rho_{halo}$ is the density profile of dark matter in the Milky
Way, $\left\langle
\int_{los}\rho_{halo}(\vec{\ell})d\vec{\ell}\right\rangle $ is the
average of the integration along the line of sight (los). In our
calculation, we adopt the Navarro-Frenk-White (NFW) halo
profile\ \cite{Navarro:1995iw}
\begin{equation}
\rho(r)=\frac{\rho_{0}}{(r/r_{c})(1+r/r_{c})^{2}},
\label{eq:profile}
\end{equation}
where $r$ is the distance from the center of Milky Way, $r_{c}=20$
kpc, and $\rho_{0}$ is set in such a way that the dark matter density
in the solar system satisfies $\rho(r_{\odot})=0.30\:{\rm
  GeV/cm}^{3}$\ \cite{Bergstrom:1997fj} with $r_{\odot}=8.5$ kpc being
the distance from the Sun to the Galactic Center.  For the background,
we use a power-law form adopted in Ref.\ \cite{Ishiwata:2008cu} 
\begin{equation}
\left[E^{2}\frac{dJ_{\gamma}}{dE}\right]_{bg}\simeq5.18\times10^{-7}E^{-0.499}\quad{\rm GeVcm}^{-2}{\rm sr}^{-1}{\rm sec}^{-1},
\label{eq:gamma_bg}
\end{equation}
where $E$ is in units of GeV.  The predicted gamma-ray flux from the
decaying $\tilde{\nu}_{R}$ are shown in Fig.~\ref{fig:gamma} together
with experimental data. Since no hadronic decay mode exists and the
amount of $\tau$ is insignificant, the predicted gamma-ray flux is
below the EGRET data for the parameters we have chosen. However, we do not regard this tension as a
serious conflict, and  we expect that the upcoming data from FGST should be
able to further probe such a prediction from the right-handed sneutrino
dark matter decay. Also, we should emphasize that the relatively
suppressed gamma-ray flux shown in Fig.~\ref{fig:gamma} is due to the
simplification we have adopted. The gamma-ray flux may fit the EGRET
data if we allow $\mu_{2,3}$ to be non-zero and/or the other
generations of the right-handed sneutrino to contribute to the dark
matter abundance. This is because the decay branching ratios of $\tau$
will increase and $q\bar{q}$ decay channel will open due to the
mixing between the right-handed sneutrinos and the Higgs bosons, then
obviously, more $\pi^0$ will be generated.

\subsection{Antiproton flux}
\label{sec:anti-proton}
As we have mentioned before, the predicted antiprotons from decay of
gravitino dark matter~\cite{Ibarra:2007wg} tend to have a tension with
the observational data.  Of course, this apparent tension may not be
serious at all, considering the large uncertainties on both
experimental data and diffusion models.  However, we should better
have a scenario in which the antiproton flux could be suppressed. In
our set-up, the right-handed sneutrino does not decay into quarks (at
least the decay rate is suppressed), and therefore the antiproton
flux is negligible. This feature will be lost once we allow
$\theta_{13}$ and/or $\mu_{2,3}$ to deviate from zero.  Or if
$\tilde{\nu}_{R1,2}$ also contribute to the dark matter abundance,
their decay generically produces the antiproton, since they can mix
with the Higgs bosons as well as the left-handed sneutrinos.  The
antiproton flux in generic parameter space will depend on those
parameters as well as the neutrino mass spectrum, and it is beyond the
scope of this paper to survey all the possibilities. However, it is
important to keep in mind that there is a set of parameters where the
antiproton production becomes negligible.

\subsection{Neutrino flux}
\label{sec:neutrino}
The main decay channel of the $\tilde{\nu}_{R3}$ is $\nu \bar \nu$ as
can be seen from Table~\ref{tab:Brs}. It has been recently studied in
detail whether the neutrinos produced by the gravitino decay can be
detected by the current and future observations~\cite{Covi:2008jy}.
In our scenario, the neutrino flux is enhanced compared to the
positron and the gamma-ray ones, and therefore the neutrino signal 
from the $\nuR$ decay can be larger by an order of magnitude compared to the gravitino of
the same mass.  Thus, the neutrino signal, especially $\nu_\tau$, may
reach the sensitivity of future experiments, e.g. Hyper-Kamiokande.  However, we need to
mention that the enhancement in the neutrino production will be lost
for another choice of the parameters.

\section{Cosmological production of $\nuR$}
\label{sec:production} 
We would like to briefly discuss here how the right-handed sneutrino
is produced in the early universe. In order to account for the
observed dark matter abundance, the cosmological abundance of $\nuR$
must satisfy~\cite{Komatsu:2008hk}
\beq
\Omega_{\nuR}h^2 +\Omega_{\tilde{\nu}_R^c}h^2\;=\;0.1099 \pm 0.0062,
\eeq
or equivalently,
\beq
\frac{n_{\nuR}+n_{\tilde{\nu}_R^c}}{s} \;\simeq\;  4.2\times10^{-12} \lrfp{m_{\nuR}}{100{\rm GeV}}{-1},
\label{abundance}
\eeq
where $n_{\nuR}$ is the number density of $\nuR$, and $s$ the entropy density
of the universe. If not only the lightest but also heavier
right-handed sneutrinos are stable in a cosmological time,
we need to sum over those right-handed sneutrinos in Eq.~\REF{abundance}.

In the case of the neutralino dark matter, it is usually assumed that
the neutralinos are thermally produced at a temperature above the
freeze-out temperature.  However, since the neutrino Yukawa coupling
$y^{\nu}$ is very small ($\sim 10^{-13}$) in our scenario, the $\nuR$
is never in equilibrium unless it has other unknown strong
interactions with the radiation in the early universe. Therefore we
need to consider non-thermal production of $\nuR$. There are several
possibilities~\cite{Asaka:2005cn}, and we consider one by one as
follows.

First of all, $\nuR$ is produced by the decay of the lightest SUSY
particle in the MSSM sector (MSSM-LSP) through the neutrino Yukawa
coupling(s).  In the presence of $R$-parity violating interactions,
the MSSM-LSP also decays into the SM particles without producing
$\nuR$. If the $R$-parity violating coupling is much larger than the
neutrino Yukawa coupling, the production of $\nuR$ will get
suppressed.  This is the case in our set-up, since a relatively large $R$-parity violation
is needed to compensate the suppression of the decay rate into the charged leptons. 
However, for another choice
of the model parameters,  the magnitude of the
$R$-parity violation tends to be as small as the neutrino Yukawa coupling, 
and the MSSM-LSP decay into $\nuR$ via the
$R$-parity conserving interactions may occur at a non-negligible rate.
This would be indeed the case if the neutrino masses are degenerate,
since the neutrino Yukawa coupling will become larger than that in the
case of normal hierarchy, and the needed coupling strength of the
$R$-parity violation to account for the positron excess becomes
correspondingly smaller.  Then some amount of $\nuR$ will be produced
by the decay of the MSSM-LSP.  In the $R$-parity conserving case, the
production from the MSSM-LSP was studied in
Ref.~\cite{Asaka:2005cn}. We can estimate the abundance in a similar
way, taking account of the suppression due to the presence of the
decay via $R$-parity violating interactions. That is to say,
contribution to the density parameter of $\nuR$ from the decay of the
MSSM-LSP, $\chi$, is given by
\beq
\Delta \Omega_{\nuR} \;=\; B_{\nu}\ \lrf{m_{\nuR}}{m_{\chi}} \Omega_\chi,
\eeq
where $B_{\nu}$ denotes the branching ratio of the $\chi$ decay
into $\nuR$, and $m_\chi$ is the mass of $\chi$.  Note that, due to
the presence of the $R$-parity violation, the $\chi$ decays faster
than the case without it, which makes it easier to satisfy the big bang
nucleosynthesis (BBN) bound on the MSSM-LSP decay.  However, it also
means that larger abundance of the MSSM-LSP is needed to produce a
right amount of $\nuR$.  In the normal hierarchy case, this would push
up the soft masses of the SUSY particles, and the naturalness may be
called into question.  On the other hand, in the case of degenerate
neutrino masses, the effect of the $R$-parity violation is relatively
smaller, and we expect that a right amount of $\nuR$ can be generated
by the MSSM-LSP decay.

Another interesting possibility is that the gravitino decay produces
$\nuR$.  The gravitinos are generated by particle scattering in
thermal plasma~\cite{Weinberg:zq,Krauss:1983ik}, and they are also
generically produced by the inflaton
decay~\cite{Kawasaki:2006gs,Asaka:2006bv,Endo:2006tf,Endo:2006qk,Endo:2007ih}.
For the gravitino mass is of ${\cal O}(100)$\,GeV, it typically decays
into the SM particles and their superpartners during BBN. The
energetic decay products can significantly change abundances of light
elements, thus the gravitino abundance is tightly constrained by BBN.
If the gravitino is the next-to-lightest SUSY particle (NLSP),
however, it will mainly decay into $\nuR$ and a neutrino, and
therefore, the BBN constraint will be greatly relaxed.  To account for
the abundance Eq.~\REF{abundance}, the reheating temperature
should be as high as about $10^{10}$\,GeV if the gravitino is mainly
thermally produced.  Note that the production of $\nuR$ from the
gravitino decay can be concomitant with the production from the
MSSM-LSP decay as well as the following mechanisms, so the reheating
temperature can be lower.

The inflaton decay may also directly produce $\nuR$. Indeed, if the
inflaton acquires a non-vanishing vev at the potential minimum, it
couples to all the matter fields as well as the gauge fields in
supergravity~\cite{Endo:2006tf,Endo:2006qk,Endo:2007ih}. However, the
smallness of the neutrino Yukawa coupling makes the branching ratio of
the decay into $\nuR$ extremely small. It is likely that we need to
assume relatively strong interactions between $\nuR$ and the inflaton
sector (or other fields whose energy dominates the energy density of
the universe after inflation) for the enhancement of contribution from
inflaton decay.

The last possibility is the scalar condensation of $\nuR$. If the position of $\nuR$
is deviated from the origin during inflation~\footnote{
If $\nuR$ is light during inflation, 
such a displacement is generically expected to arise from the quantum fluctuations of the $\nuR$.
The fluctuations will become isocurvature fluctuations in the CDM. It is also possible to
generate large amount of non-Gaussianity in the CDM isocurvature perturbations~\cite{Kawasaki:2008sn}. 
}, $\nuR$ will start oscillating after inflation when
the Hubble parameter becomes comparable to its mass. The abundance is estimated to be
\beq
\frac{n_{\nuR}}{s} \;\simeq\; \frac{T_R}{4 m_{\nuR}} \lrfp{\tilde{\nu}_{R}^{osc}}{M_P}{2} 
\sim 2 \times 10^{-13} \lrfp{m_{\nuR}}{100{\rm GeV}}{-1}
\lrf{T_R}{10^{6}{\rm GeV}} \lrfp{\tilde{\nu}_{R}^{osc}}{10^{10}{\rm GeV}}{2},
\eeq
where $T_R$ denotes the reheating temperature of the universe and
$\tilde{\nu}_{R}^{osc}$ is the amplitude of the oscillations.  We have
assumed that the sneutrino starts to oscillate before the reheating is
completed. For an appropriate choice of the amplitude, it can account
for the right abundance of the observed dark matter. Note that the
$\Omega_{\nuR}$ is determined solely by the reheating temperature and
the initial position of $\tilde{\nu}_R$ in this case, and is independent of the sneutrino
mass.

\section{Discussion and Conclusions}
\label{sec:Conclusion} 
We have assumed that the $R$-parity is dominantly violated by the
bilinear term. If it is violated mainly by the trilinear terms,
$LL \bar{E}$, $Q\bar{D}L$, or $\bar{U} \bar{D} \bar{D}$, the predicted cosmic-ray
spectra will be different. If the $R$-parity is broken by the $LL \bar{E}$
operator, the $\nuR$ will decay into charged leptons as well as 
neutrinos through the left-right mixing, which will result in a sharp
peak in the predicted positron fraction at the solar system. The
continuum gamma rays are also produced, while
virtually no antiprotons are produced. This choice therefore provides an ideal 
way to avoid the observational constraint on the antiproton flux. In the case of $Q\bar{D}L$ or
$\bar{U} \bar{D} \bar{D}$, the hadronic branching ratios are larger than the case of $LL \bar{E}$, 
and therefore we expect a relatively large
contribution to the antiproton flux.

With the tiny $R$-parity violations in our scenario, the NLSP is
generally long-lived in collider experiments and decays outside the
detector~\cite{Hamaguchi:2004df,Hamaguchi:2006vu}.  If NLSP is the neutralino, it is observed as a missing
energy, and the collider phenomenology is the same as the $R$-parity
conserved case. If stau is the NLSP, we expect to observe its track
inside the collider, which is similar to the case of the decaying gravitino
LSP~\cite{Ishiwata:2008cu}. However, there is one possibility that the $\nuR$ is only a fraction
of the total dark matter, and the rest is explained by some other
stable particles, such as a QCD axion. If the fraction of $\nuR$ in dark
matter $r$ is very small, the $R$-parity violating coupling should be
enhanced as $\propto r^{-1/2}$, for a fixed positron flux. Since the
lifetime must be longer than the present age of the universe, $r$
can be as small as ${\cal O}(10^{-10})$, and the corresponding $R$-parity
violation can be enhanced by a factor of  ${\cal O}(10^{5})$. With such a large enhancement in the
 $R$-parity violation, we may
observe some signatures of decaying NLSP at colliders. Note also that, even if the
fraction is much smaller than unity, the features of the cosmic-ray
spectra from decaying right-handed sneutrino  will not
change.

We have focused on the case that all the right-handed sneutrinos have
almost the same mass, and considered only one of them is the source of
cosmic-ray. If the three right-handed sneutrinos exist with an equal
amount in our universe, they will all contribute to the cosmic-ray.
The contribution from one of the three right-handed sneutrinos may dominate over those from
the other two, depending on the values of $\mu_i$ and the neutrino mass
spectrum.

Even there exist mass differences among the three right-handed
sneutrinos, unless the decay channel for the heavier one to a on-shell
higgsino is kinematically allowed, the decay patterns of all the
  right-handed sneutrinos are similar to each other. Namely, the
heavier $\nuR$ also decays via $R$-parity violating interactions,
since the $R$-parity conserving decay channel is suppressed as
explained in the following paragraph.

We should also emphasize here that the excess in the cosmic-ray
positron can be also explained without introducing the $R$-parity
violation interactions.  If the right-handed sneutrinos are the three
lightest supersymmetric particles, and if there is slight hierarchy in
the masses, the heavier one will decay into a lighter one accompanied
by two SM leptons through an off-shell higgsino. It is obvious that
the two SM leptons will be the source for the cosmic positron and
gamma-ray excesses. Also, the antiproton will not be
produced. However, the lifetime tends to be much longer than the needed one
to account for the positron excess, since the decay amplitude is
suppressed by the neutrino Yukawa coupling squared. One solution is to
add a small Majorana mass term for the right-handed
neutrinos~\cite{Gopalakrishna:2006kr}. Then the see-saw mechanism will
occur at very low energy, and the neutrino Yukawa coupling becomes
larger, which can make the lifetime of the heavier one short enough to
explain the positron excess.  Another way to make the lifetime shorter
is to tune the higgsino mass close to the heavier right-handed sneutrino mass.

It is even possible to explain the positron excesses in a
non-supersymmetric theory.  One candidate is the sterile
neutrino dark matter, whose decay produces charged leptons and neutrinos.
The decay processes of a sterile neutrino lighter than $100$\,MeV
were studied in a different context (see e.g. \cite{Picciotto:2004rp}).  
Since the sterile neutrino must be heavier than $100$\,GeV in our case,
it will decay also into $W$ and $Z$ bosons, and also into $\nu \gamma$
with a suppressed rate. Thus, the energy spectra of the cosmic-ray particles
from the sterile neutrino decay will look similar to those in the case of the decaying 
gravitino. The mixing angle between the sterile and active neutrinos must be extremely suppressed
in order to realize the long lifetime of the sterile neutrino, which
will make it difficult to produce the sterile neutrino through the
mixing. It will be necessary to introduce a non-thermal production
mechanism of the sterile neutrino.

So far, only the normal hierarchy case in neutrino mass spectrum is
considered, but we expect that an essential feature of our results
will persist in the other neutrino mass spectra.  Rather, it may be
possible to extract some information on the neutrino mass spectrum
from the positron and gamma-ray fluxes by the
PAMELA and FGST in operation. We leave those issues for future work.

We propose a leptonic dark matter candidate whose decay may explain the excesses of observed 
cosmic rays. If the dark matter has a
lepton number, it will not be surprising that we see some excesses
only in the positron and the gamma-ray fluxes while virtually no
primary antiprotons are observed.  To illustrate the idea, in this
paper we have studied a scenario that the neutrinos are Dirac type,
and the right-handed sneutrino accounts for the observed dark matter
and decays into the SM particles through the $R$-parity violating
interactions.  The charged leptons as well as neutrinos are directly
produced during the decay, leading to a sharp rise in the positron
fraction.  The decay products will also generate continuum gamma rays,
which may account for the EGRET excess. Interestingly, with an
appropriate set of the parameters, the antiproton production can be
suppressed in our scenario, which is consistent with an observation
that most of the observed antiprotons are secondaries. Furthermore,
we expect that the energy spectra for those
cosmic-ray particles may provide a new probe into the neutrino mass
spectrum. Those features on the predicted cosmic-ray spectra in our
study will be checked by the PAMELA and FGST satellites.

\bigskip 
{\it Note added}:
After our paper was posted, the PAMELA group reported a steep rise in
the positron fraction~\cite{Adriani:2008zr}, which can be explained by
our model with a 300 GeV right-handed sneutrino dark matter,
 as shown in Fig.~\ref{fig:Fraction_e}.

\begin{acknowledgements} This work was supported by World Premier
International Research Center Initiative (WPI Initiative), MEXT, Japan.
\end{acknowledgements}




\end{document}